\documentclass[prb,twocolumn,showpacs,preprintnumbers]{revtex4}
\usepackage{graphics}
\usepackage{graphicx}
\usepackage{dcolumn} 
\usepackage{bm}
\usepackage{float}
\usepackage{epsfig}
\begin{document}
\title{On kinetic energy stabilized superconductivity in cuprates}
\author{D.J. Singh}
\affiliation{Materials Science and Technology Division,
Oak Ridge National Laboratory, Oak Ridge,TN 37831-6032}
\date{\today}
\pacs{}

\begin{abstract}
The possibility of kinetic energy driven superconductivity in
cuprates as was recently found in the $tJ$ model is discussed.
We argue that the violation of the virial theorem implied by this
result is serious and means that the description of superconductivity
within the $tJ$ model is pathological.
\end{abstract}

\maketitle

Recent numerical simulations and analytical studies of the
Hubbard and $tJ$ models,
which suggest that superconductivity in doped cuprates
may be driven by a lowering of kinetic energy upon formation of
the superconducting state
\cite{wrobel1,wrobel2,eckl,feng,yokoyama,maier1,maier2,yanase,ogata,gu,lee}
have resulted in considerable interest,
including spectroscopic measurements and their interpretation.
\cite{santander,molegraaf,kuzmenko}
The essential physics depends on the replacement of the full 
Hamiltonian and electronic degrees of freedom by a limited
number of degrees of freedom associated with the charges of the
doped carriers within a stiff background related to the electronic
structure of the undoped antiferromagnetic Mott insulating cuprate phases,
which motivate these models. Within this antiferromagnetic
background paired carriers
can be more mobile than single carriers, and this can overcome
the normal increase in kinetic energy upon pair formation.
As was already mentioned, such an increase in kinetic energy is
unusual as normally the virial theorem of Clausius would prevent
it, but it has been argued that
this is allowed within the $tJ$ model,
and that the kinetic energy as defined within the $tJ$
model is a more physical quantity than the kinetic energy
of the full electronic system. \cite{wrobel1}

The purpose of this short paper is to argue that this apparent
violation of the virial theorem is serious, and implies that if
the result that superconductivity in the $tJ$ model is kinetic
energy driven is correct, then the $tJ$ model is pathological in
the sense that its superconductivity is essentially different from
that in the cuprates, or that it neglects degrees of freedom that
are essential for obtaining the superconducting state, or both.

To a very good approximation, solids, including cuprates, may be
regarded as bound systems composed of nuclei and electrons interacting
via the Coulomb potential. While highly precise, this view is generally
useless without approximations and effective models that capture
the physics of interest and the relevant degrees of freedom within
a tractable framework.
Constructing such models is crucial for progress, not only because
they make calculations possible, but also because by reducing
the number of degrees of freedom to a smaller number of approximate
degrees of freedom that can be physically interpreted, they provide
understanding of the essential physics.
However, exact results and scalings, based
on the bare system are useful in constraining effective models
and defining their range of applicability. For example, the use
of exact scalings for the electron gas has proved to be of
considerable value in
the construction of generalized gradient approximations for
density functional studies of solids. \cite{perdew,perdew2}

According to the virial theorem for a stable bound system of interacting
Coulomb particles, \cite{clausius,fock,slater}
the kinetic energy, $T$, which is positive, is
equal to -1/2 of the potential energy, $V$, which is negative
with standard definitions. Thus the
energies of eigenstates are ordered such that lower energy corresponds to
higher kinetic energy. This applies also to a variety of other local
potentials and to relativistic systems.
Since the superconducting state in cuprates is a ground state, or if not,
it is at least a lower energy state than the ensemble that comprises the
normal state at finite temperature (the specific heat is positive),
the virial theorem implies that the conventional
kinetic energy in the superconducting
state is unambiguously higher than that in the normal state.

Therefore, in cuprates, using conventional definitions of potential
and kinetic energy, superconductivity is driven by a reduction in
potential energy, accompanied by a smaller increase in kinetic energy.
Norman and co-workers \cite{norman}
have discussed the relationship between the condensation energy
and electronic spectral functions, especially as related to angle resolved
photoemission. They emphasize that while the virial theorem applies
to the full Hamiltonian, it need not apply in a reduced subspace, e.g.
the space of low energy electronic excitations.
In any case,
a simple interpretation of the above would be that the $tJ$ model
describes an unphysical superconductivity. However, the situation
may not be so simple. First of all, as was suggested already, \cite{wrobel1}
the kinetic energy may be redefined as the kinetic energy of the
lower Hubbard band, and this is essentially the quantity that decreases
in the $tJ$ model. Secondly,
it may be argued that the kinetic energy
overall increases, but that the kinetic energy relevant for excitations
up to some cut-off energy decreases.
However, both of these scenarios, which are related, would
require that there be a larger kinetic energy increase
involving degrees of freedom not included in the $tJ$ model.
This means that the main physics driving superconductivity is not
in the $tJ$ model, or that the kinetic energy of the $tJ$ model is to
be interpreted as
mainly potential energy of the bare Hamiltonian, which
would be difficult to understand since it originates in the hopping
term.

We now turn to the origin of the virial theorem violation in the $tJ$
model and speculate about possible ways forward.
As mentioned, the kinetic energy decrease into the superconducting
state of the $tJ$ model is apparently connected with the stiff
antiferromagnetic Mott insulating background into which a small
number of carriers are doped in this view of cuprate superconductivity.
However, while the phase diagrams of cuprate superconductors generally
show prominent Mott insulating phases at zero doping, these are
separated from the superconducting phases. The superconducting
transition is between a high temperature conducting state, with
specific heat and other thermodynamic properties similar to a high
carrier density metal and a unconventional superconductor.
In general, the Fermi surfaces in the normal state, as measured by
a variety of probes, are consistent with a high carrier density.
\cite{pickett}
The transport, on the other hand, especially in the underdoped regime,
shows a variety of non-Fermi liquid scalings, for example, linear in $T$
resistivity. The non-Fermi liquid scalings evolve continuously into
conventional metallic behavior with increasing hole doping above the
optimum for $T_c$.
These unconventional
scalings can be reproduced within the framework of strong
correlated models, based on doping of an underlying Mott insulating state.
\cite{lee}
However, it should be noted that the Mott insulator -- metal
transition is thought to be first order in clean cuprates,
and therefore the connections between the Mott insulating phase
and the conducting phase need not be taken for granted.
Non-Fermi liquid scalings occur in other correlated
materials that cannot be regarded as doped Mott insulators, for
example, metals near quantum critical points.
\cite{hertz,lonzarich}

The importance of retaining charge degrees of freedom, not included in
the $tJ$ model has also been discussed as by Phillips and co-workers
\cite{phillips1,phillips2} both
from the point of view of assymptotic freedom and from
the metallic character.

The spin-charge separation that occurs in the $tJ$ model and is expected
in lightly doped Mott insulators, is not essential for non-Fermi liquid
scalings in transport nor is the stiff antiferromagnetic background.
Soft fluctuations that scatter charge carriers would suffice.
In fact, based on neutron scattering experiments \cite{keimer,aeppli,birgeneau}
and analysis of
nuclear magnetic resonance (NMR) data, \cite{monthoux}
the antiferromagnetic correlation
length strongly decreases with doping and is $\sim$ 3 lattice spacings
or less at optimal doping.
This may be important both for the normal state properties
and the superconductivity, although we note that a theory for
cuprate superconductivity has not yet been established.
A scenario in which doped cuprates are high carrier density metals
with strong $T$-dependent scattering due to a nearby quantum critical
point is more likely if the soft quantum fluctuations have a short
coherence length, than if they are due to a sharply peaked (in $Q$)
structure.
This is because in the former case there is a larger phase
space for fluctuations, and therefore a stronger quantum suppression of
the underlying instability and more scattering. This can be seen from
the fluctuation dissipation theorem, which relates the imaginary part
of the susceptibility and the fluctuation amplitude.
\cite{moriya1,ggl,moriya2,moriya3}
Furthermore, within a spin-fluctuation
pairing Migdal-Eliashberg framework, fluctuations with a short coherence
length such that the inverse coherence length is of the order of the
reciprocal space length scale for variation of the $d_{x^2-y^2}$
order parameter on the Fermi surface (i.e. $\sim$ 3-4 lattice spacings),
would be more effective for pairing than sharply peaked fluctuations at the
antiferromagnetic wavevector.
\cite{monthoux,moriya3}
This is again because of phase space arguments, specifically that
increasing the coupling at a specific $k$ leads to ordered antiferromagnetism,
while increasing the range of $k$ involved strengthens the overall
pairing without producing a magnetic instability.

In any case, assuming that calculations showing kinetic energy driven
superconductivity in the $tJ$ model for cuprates are correct, we
argue that the $tJ$ model is insufficient for understanding superconductivity
in cuprates. One avenue for going forward may be to add more degrees of
freedom in extended models to produce a softer, more metallic normal state
especially in the charge channel. By removing the stiff nearly
antiferromagnetic Mott insulating background, this may destroy the
artificial kinetic energy driven $tJ$ model
superconductor in favor of a superconducting
state consistent with the virial theorem.

I am grateful for helpful discussions with R.S. Fishman, D.I. Khomskii
and M.R. Norman.
Research at ORNL
sponsored by the Division of Materials Sciences and Engineering,
Office of Basic Energy Sciences, U.S. Department of Energy,
under contract DE-AC05-00OR22725 with Oak Ridge National Laboratory,
managed and operated by UT-Battelle, LLC.

\end{document}